\documentclass{iauJDSS} 

\usepackage{bm}
\usepackage{amssymb}
\usepackage{amsmath}

\def\be{\begin{equation}}
\def\ee{\end{equation}}
\def\ba{\begin{eqnarray}}
\def\ea{\end{eqnarray}}

\def\a{\alpha}
\def\b{\beta}

\def\m{\mu}
\def\n{\nu}

\title[Celestial dynamics and astrometry in expanding universe]{Celestial dynamics and astrometry in expanding universe} 
\author{Sergei Kopeikin}
\affiliation{Department of Physics \& Astronomy, University of Missouri, 322 Physics Bldg.,\\ Columbia, MO 65211, USA}
\pubyear{2012}
\volume{Volume 15}  
\pagerange{???--???}
\date{}
\jname{Highlights of Astronomy, Volume 16}
\editors{N. Capitaine, G. Kaplan \& S. Klioner, eds.}
\begin{document}
\maketitle
\abstract{The mathematical concept of the Newtonian limit of Einstein's field equations in the expanding Friedmann universe is formulated. The geodesic equations of motion of planets and light are derived and compared.}

The current paradigm of the IAU 2000 resolutions assumes that the background spacetime is flat with the Minkowski metric tensor $\eta_{\a\b}$. A more adequate model is the Friedmann spacetime with the metric $\bar g_{\a\b}=a^2(\eta)\eta_{\a\b}$, where $a(\eta)$ is the scale factor describing the Hubble expansion. The coordinate chart is $x^\a=(c\eta,x^i)$, where the conformal (non-physical) time $\eta$ relates to the cosmic (physical) time $t$ by differential equation $a(\eta)d\eta=dt$. Time $t$ is the proper time measured by the Hubble observers. We assume that $\eta=0$ corresponds to the present epoch.

The solar system is a localized astronomical system which gravitational field exceeds the strength of the gravitational field of the universe by many orders of magnitude. It makes sense to introduce the {\it local} coordinates, $r^\a=(c\lambda,r^i)$, that are obtained from the {\it global} coordinates, $x^\a$, by a {\it special conformal} transformation, $r^\a=(x^\a-b^\a x^2)/\sigma(x)$, where $b^\a$ is a constant vector, $\sigma(x)=1-2b_\m x^\m+b^2(x_\n x^\n)$, and $
x^2\equiv \eta_{\m\n}x_\m x^\n$, $b^2\equiv \eta_{\m\n}b_\m b^\n$ [\cite{1}]. The vector $b^\a$ is chosen so that, $a^2(\eta)\eta_{\a\b}dx^\a dx^\b=\eta_{\a\b}dr^\a dr^\b$, is a valid equation up to the second order in the Hubble constant, ${\cal H}= a'/a$ (a prime denotes a time derivative w.r.t. $\eta$). Transformation to the local metric is achieved with $b^\a=-(1/2c){\cal H}\bar u^\a$, where $\bar u^\a$ is four-velocity of the Hubble flow. Approximating $a(\eta)=1+{\cal H}\eta$ yields, $\lambda=a(\eta)[\eta-(1/2){\cal H}(c^2\eta^2-{\bm x}^2)]$, and ${\bm r}=a(\eta){\bm x}$.

We have to relate the local time $\lambda$ to the fundamental cosmological time $t$. In the linearized approximation $\eta=t-(1/2){H}t^2$, where ${H}={\cal H}/a$ is referred to time $t$. For particle being at rest, the conformal transformation of time yields $\lambda=t+const.$ The same relationship is hold for slowly moving particles but light moves along null geodesics, $c^2\eta^2-{\bm x}^2=0$. The conformal time transformation yields the relation of parameter $\lambda$ of the light geodesics with the cosmic time, $\lambda=t+(1/2){H}t^2$. It means that in the Newtonian-limit the metric for slowly-moving particles is $\eta_{\a\b}$ but the optical metric for light is, $\hat g_{\a\b}=\eta_{\a\b}+(1-a^2)\bar u_\a\bar u_\b$, and explicitly depends on the scale factor, $a=a(t)$. This is interpreted as a non-dispersive index of refraction, $n=1/a(t)$, that makes the coordinate speed of light, $c_l=c/n=a(t)c$, growing as time goes from past to future, and decreases otherwise.

The equations of motion of particles and light are reduced to their Newtonian form for both metrics. However, equations for particles and for light differ by terms of the first order in the Hubble constant [\cite{2}]. This leads to the important conclusion that the equations of motion for propagation of light used currently by Space Navigation Centers for calculating ephemerides of planets and spacecrafts, are missing some terms of cosmological origin. The correct light-propagation equation is, $\lambda_2-\lambda_1=(1/c)|{\bm r}_2-{\bm r}_1|$, where $\lambda=t+(1/2){H}t^2$. Inclusion of the quadratic-in-time terms to the data processing of Doppler tracking eliminates the, so-called, Pioneer anomaly [\cite{3}] which may really have a cosmological origin. 

\thebibliography{10}
\bibitem[1]{1}{Fulton, T., Rohrlich, F. \& Witten, L.} 1962, \textit{Rev. Mod. Phys.}, 34, 442
\bibitem[2]{2}{Kopeikin, S.} 2012, \textit{PRD}, 86, 064004
\bibitem[3]{3} {Anderson, J D., Laing, P. A., Lau, E. L., Liu, A. S., Nieto, M. M. \& Turyshev, S. G.} 2002, \textit{PRD}, 65, 082004
\end{document}